\documentclass[10pt]{article}
\usepackage{opex2}
\begin{document}

   \title{Instabilities of a Bose-Einstein condensate in a periodic
potential: an experimental investigation}
   \author{M.~Cristiani, O.~Morsch, N.~Malossi, M.~Jona-Lasinio, M.~Anderlini, E.~Courtade and
E.~Arimondo}\address{Istituto Nazionale per la Fisica della
Materia, Dipartimento di Fisica E. Fermi, Universit\`{a}
    di Pisa, Via Buonarroti 2, I-56127 Pisa,Italy}
   \date{\today}
   \begin{abstract}
   By accelerating a Bose-Einstein condensate in a controlled way
   across the edge of the Brillouin zone of a 1D optical lattice,
   we investigate the stability of the condensate in the vicinity
   of the zone edge. Through an analysis of the visibility of the
   interference pattern after a time-of-flight and the widths of
   the interference peaks, we characterize the onset of
   instability as the acceleration of the lattice is decreased. We
   briefly discuss the significance of our results with respect to
   recent theoretical work.
   \end{abstract}

   \section{Introduction}
   Bose-Einstein condensates (BECs) trapped in the periodic
potentials created by
   optical lattices have become a thriving field of research in
   recent years~\cite{menotti03,wu03,louis03}. Since such systems are
very similar in their
   theoretical description to electrons in a solid state crystal,
   many experiments have been carried out demonstrating phenomena
   that were, in some cases, well-known from condensed matter
   physics, e.g. Bloch oscillations and Landau-Zener
tunneling~\cite{morsch01,cristiani02}. In a recent experiment,
   a BEC in an optical lattice even made possible the observation
   of a quantum phase transition that had, up to then, only been
   theoretically predicted for condensed matter
   systems~\cite{greiner02}.

   Apart from the latter, most
   experiments to date have been carried out in the regime of
   shallow lattice depth, for which the system is well described
   by the Gross-Pitaevskii equation with a periodic potential. The
   nonlinearity induced by the mean-field of the condensate is
   included in this description and has been shown both theoretically
   and experimentally to give rise to
instabilities~\cite{wu03,cataliotti03,kivshar94,konotop02,hilligsoe02,scott03,anglin03}
in certain regions of the Brillouin zone. In
   this paper we experimentally study these instabilities by
   accelerating the optical lattice and thus scanning the
   Brillouin zone in a controlled way.

Nonlinearity-induced instabilities are observed in many different
branches of physics, and they provide a dramatic manifestation of
strongly nonlinear effects (see review paper \cite{kivshar00}).
The effects of modulational and transverse instabilities have been
observed mostly for continuous media. However, periodic structures
can strongly influence such instabilities.  The growth of the
instability can be controlled in periodic structures where the
effective geometric dispersion provides a key physical mechanism
for manipulating waves in various physical systems, including
Bragg gratings in optical fibers, waveguide arrays and, more
recently, optically induced photonic lattices and crystals, as
seen in recent observations of the propagation characteristics of
a probe laser beam~\cite{fleischer03,sukhorukov03,neshev03}.

In this Letter we study, both theoretically and experimentally,
matter wave propagation in optically-induced lattices near the
condition for Bragg reflection. Formally, these experiments are
closely linked to light scattering in periodic structures and can
be described by similar mathematical models. In particular, our
experiments reveal the role of nonlinear Bloch-wave interactions
in periodic media.

This paper is organized as follows. After briefly describing in
   section 2 our experimental apparatus and the method
   used for scanning the Brillouin zone in a controlled way, we
   present the results of our experiments in
   section 3. A comparison between these results and a
   1-D numerical simulation is the subject of section 4,
   followed by a discussion of our findings and prospects for
   future investigations (section 5).
   \section{Experimental setup and procedure}\label{setup}
   Our experimental apparatus for creating BECs of ${^{87}}Rb$ atoms
   is described in detail in~\cite{muller00}. The main feature of
   our apparatus relevant for the present work is the
   triaxial time-averaged orbiting potential (TOP) trap with
   trapping frequencies $\nu_x:\nu_y:\nu_z$ in the ratio
   $2:1:\sqrt{2}$. Our trap is, therefore, almost isotropic. The
   consequences  for our experiment of this near isotropy will be
   discussed at the end of this paper.\\
   The optical lattice is created by two counterpropagating laser
   beams with parallel linear polarizations and wavelength $\lambda$,
as described in
   detail in~\cite{morsch01,cristiani02}. The two beams are derived
from the first diffraction orders of two
   acousto-optic modulators that are phase-locked but whose
frequencies are independent, allowing us to introduce a
   frequency difference $\Delta \nu$ between them. The resulting
periodic potential has a
   lattice constant $d=\lambda /2= 0.39\,\mathrm{\mu m}$, and the
   depth of the potential (depending on the laser intensity and
detuning from the atomic resonance of the rubidium atoms)
    can be varied between $0$ and
   $\approx 2\,E_{rec}$ ($E_{rec}=\frac{\hbar^2 \pi^2}{2md^2}$). In
addition, by linearly chirping the frequency
   difference $\Delta \nu$, the lattice can be accelerated with
   $a=d\frac{d\Delta\nu}{dt}$. In our experiments, we used
   accelerations ranging from $a=0.3\,\mathrm{m\,s^{-2}}$ to
   $a=5\,\mathrm{m\,s^{-2}}$.\\
   The experimental protocol for `moving' the condensate across the
Brillouin zone is as
   follows. After creating BECs with $\approx 10^4$ atoms, we
   adiabatically relax the magnetic trap frequency to
   $\nu_x=42\,\mathrm{Hz}$. Thereafter,
   the intensity of the lattice beams is ramped up from $0$ to a
   value corresponding to a lattice depth of $\approx 2\,E_{rec}$.
   The ramping time is of the order of several milliseconds in
   order to ensure adiabaticity~\cite{band02}. Once the final lattice
   depth has been reached, the lattice is accelerated for a time $t$.
   Finally, both the magnetic trap and the optical lattice are
switched
   off, and the condensate is observed after a time-of-flight of
   $21\,\mathrm{ms}$ by absorption imaging.
   \begin{figure}[htbp]
\centering\begin{center}\mbox{\epsfxsize 3.0 in
\epsfbox{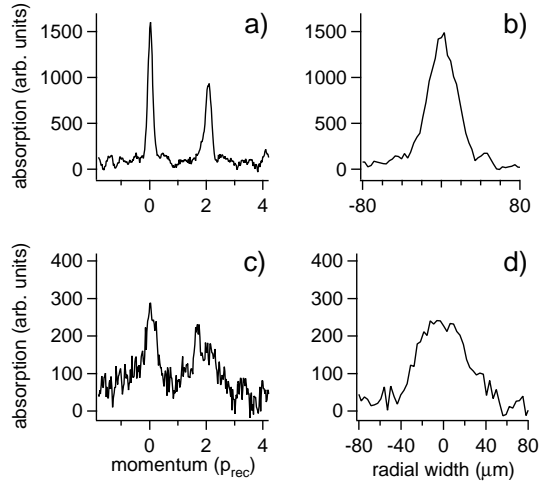}} \caption{Integrated longitudinal
and transverse profiles of the interference pattern of a
condensate released from an optical lattice after acceleration to
a quasimomentum $\approx 0.9$ and a subsequent time-of-flight of
$21\,\mathrm{ms}$. In (a) and (b), the acceleration $a$ was
$5\,\mathrm{m\,s^{-2}}$, whereas in (c) and (d)
$a=0.3\,\mathrm{m\,s^{-2}}$. In (a) and (c), the horizontal axis
has been rescaled in units of recoil momenta. Note the different
vertical axis scales (by a factor $4$) for the upper and lower
graphs. The total number of atoms was measured to be the same in
both cases.}\label{peaks}
\end{center}\end{figure}
   \section{Results: visibility and radial width}\label{results}
   When the lattice is accelerated, the condensate feels a force
   in the rest frame of the lattice, resulting in a change of
   quasimomentum of the condensate. In the linear problem, this
   simply means that when switching off the lattice and magnetic
   trap at the end of the acceleration process, the instantaneous
   group velocity of the condensate in the lattice frame is
   given by the inverse of the curvature of the lowest
   energy band of the lattice. The resulting Bloch oscillations have
been observed in a previous experiment~\cite{morsch01}.
   In practice, the time-of-flight
   interference pattern of the condensate released from the
   lattice then consists of a series of well-defined peaks
   corresponding to the momentum classes (in multiples of the
   lattice momentum $2p_{rec}=2\hbar k_L$ with $k_L=2\pi / \lambda$),
as can be
   seen in Figs.~\ref{peaks} (a) and (c). The shape of the
interference pattern in the transverse direction is shown in
Figs.~\ref{peaks} (b) and (d).

   In the nonlinear problem, the
   solutions of the Gross-Pitaevskii equation are predicted to be
   unstable in the vicinity of the Brillouin zone
edge~\cite{wu03,kivshar94,scott03,anglin03}. When the
   condensate is close to the zone edge, the unstable solutions
   grow exponentially in time, leading to a loss of phase
   coherence of the condensate along the direction of the optical
   lattice. In our experiment, the time the condensate spends in
   the `critical region' where unstable solutions exists is varied
   through the lattice acceleration. When the acceleration is
   small, the condensate moves across the Brillouin zone more
   slowly and hence the growth of the unstable modes~\cite{wu03}
becomes more important.
   Figures~\ref{peaks} (c) and (d) show typical integrated profiles
of the interference pattern for a
   lattice acceleration $a=0.3\,\mathrm{m\,s^{-2}}$. Here, the
   condensate has reached the same point close to the Brillouin
   zone edge as in Figs.~\ref{peaks} (a) and (b), but because of the
longer time
   it has spent in the unstable region, the interference pattern
   is almost completely washed out. It is also evident that the
radial expansion of
   the condensate is considerably enhanced when the Brillouin zone
   is scanned with a small acceleration.

   \begin{figure}[htbp]
\centering\begin{center}\mbox{\epsfxsize 4.0 in
\epsfbox{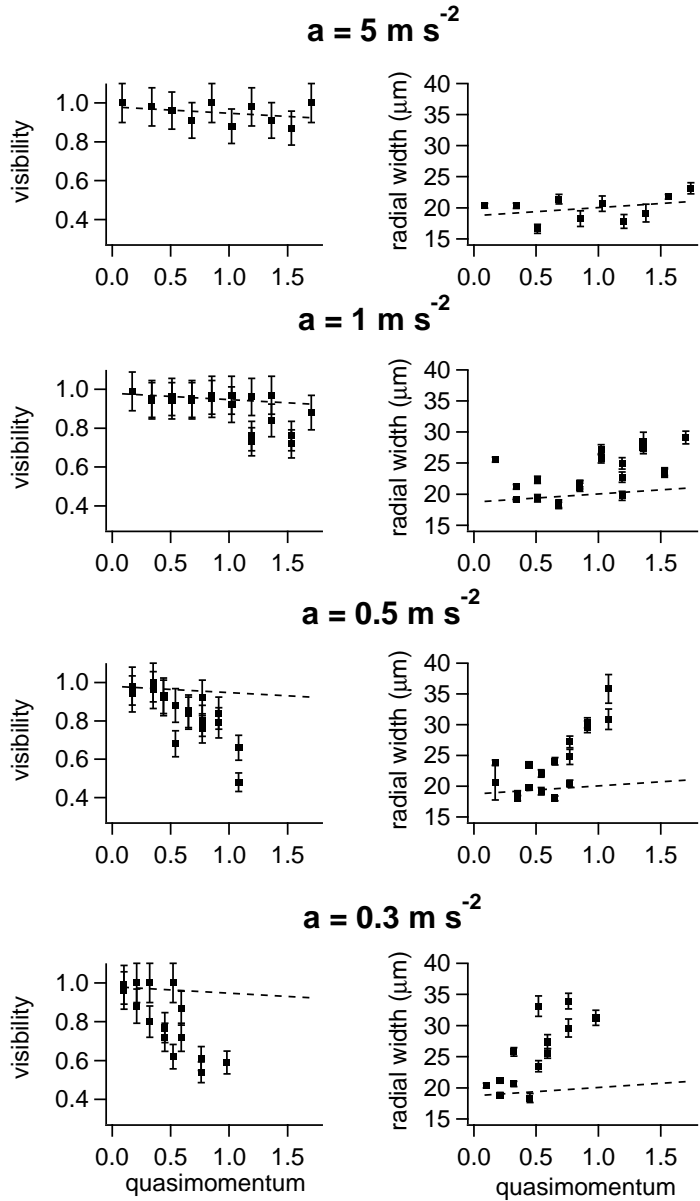}} \caption{Visibility and radial width as
a function of quasimomentum (in units of $p_{rec}$) for different
accelerations. As the acceleration is lowered, instabilities close
to quasimomentum $1$ (corresponding to the edge of the Brillouin
zone) lead to a decrease in visibility and increase in radial
width. For comparison, in each graph the (linear) fits to the
visibility and radial width for the $a=5\,\mathrm{m\,s^{-2}}$ data
are included. The error bars on the visibility correspond to an
estimated $10\%$ systematic error, whereas the error bars on the
radial width are the standard deviations of the Gaussian fits.
}\label{instability}
\end{center}\end{figure}
   In order to characterize our experimental findings more
   quantitatively, we define two observables for the
   time-of-flight interference pattern. By integrating the profile
   in a direction {\em perpendicular to} the optical lattice
direction, we obtain
   a two-peaked curve (see Fig.~\ref{peaks} (a) ) for which we can
define a visibility (in
   analogy to spectroscopy) reflecting the phase coherence of the
   condensate (visibility close to $1$ for perfect coherence,
   visibility $\longrightarrow 0$ for an incoherent condensate).
   In order to avoid large fluctuations of the visibility due to
   background noise and shot-to-shot variations of the
   interference pattern, we have found that a useful definition of
   the visibility is as follows:
   \begin{equation}
   visibility = \frac{h_{peak}-h_{middle}}{h_{peak}+h_{middle}},
   \end{equation}
   where $h_{peak}$ is the mean value of the two peaks (both averaged
over $1/10$ of their separation symmetrically about the positions of
the peaks).
   By averaging the longitudinal profile over $1/3$ of the peak
separation symmetrically about the midpoint between the peaks, we
obtain
   $h_{middle}$. For instance, applying this definition to the
profiles shown in Figs.~\ref{peaks} (a) and (c),
    we obtain $visibility=0.98$ and $visibility=0.6$, respectively.
   Owing to fluctuations in the background and hence the
   definition of the zero point of the longitudinal profile, the
   visibility thus measured can slightly exceed unity, in which case
we
   define it to be $1$.
   The second observable is the width of a Gaussian fit to the
   interference pattern integrated {\em along} the lattice direction
   over the extent of one of the peaks (see Fig.~\ref{peaks} (b) and
(d)).

   The results of our experiment are summarized in
Fig.~\ref{instability}.
   For four different accelerations ($a=5,1,0.5$ and
   $0.3\,\mathrm{m\,s^{-2}}$) we measured the visibility and
   radial width of the interference pattern as a function of the
   quasimomentum of the condensate. For large accelerations, the
   quasimomentum can be simply calculated from $a$
   and the duration of the lattice acceleration, whereas for small
   accelerations the restoring force of the magnetic trap has to
   be taken into account as the spatial motion of the condensate
   becomes appreciable. In these cases, we derived the
   quasimomentum reached in the experiment from a numerical
integration of the semi-classical
   equations of motion of the condensate in the presence of the
   periodic potential (giving rise to Bloch oscillations due to the dispersion relation of the lowest energy band) and of the magnetic trap.\\
   Figure~\ref{instability} shows clearly that for accelerations
   down to $1\,\mathrm{m\,s^{-2}}$, both the visibility and the radial
   width of the interference pattern remain reasonably stable when
   the edge of the Brillouin zone is crossed. In contrast, for
   $a=0.5\,\mathrm{m\,s^{-2}}$ and $a=0.3\,\mathrm{m\,s^{-2}}$ one
   clearly sees a drastic change in both quantities as
   the quasimomentum approaches the value $1$. For those
   accelerations, the condensate spends a sufficiently long time
   in the unstable region of the Brillouin zone and hence loses
   its phase coherence, resulting in a sharp drop of the
   visibility. At the same time, the radial width of the
   interference pattern increases by a factor between $1.5$ and
   $1.7$. This increase is evidence for an instability in the
   transverse direction and may, for instance, be due to solitons
   in the longitudinal direction decaying into vortices via a
   snake instability~\cite{scott03}. For $a=0.5\,\mathrm{m\,s^{-2}}$ and
   $a=0.3\,\mathrm{m\,s^{-2}}$, the interference patterns for
   quasimomenta larger than $1$ were so diffuse that it was not
   possible to measure the visibility nor the radial width in a meaningful way.
\begin{figure}[htbp]
\centering\begin{center}\mbox{\epsfxsize 3.0 in
\epsfbox{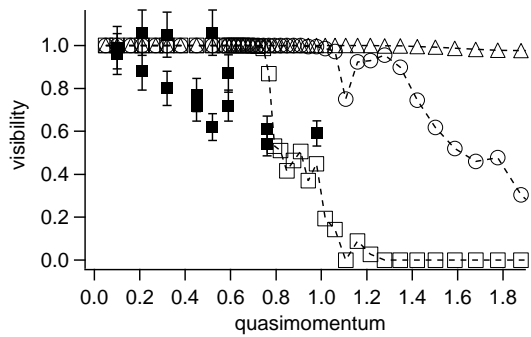}} \caption{Results of a one-dimensional
numerical simulation of our experiment for different values of the
nonlinear parameter $C$ and acceleration
$a=0.3\,\mathrm{m\,s^{-2}}$. The open squares, circles and
triangles correspond to $C=0.008$ (the value for our experiment),
$C=0.004$ and $C=0$, respectively. The closed symbols are the
experimental values of the visibility as reported in Fig.~2 for
$a=0.3\,\mathrm{m\,s^{-2}}$. The dashed lines connect the
theoretical points to guide the eye.}\label{instab_theo}
\end{center}\end{figure}

   \section{Comparison with a 1-D numerical
simulation}\label{numerics}
   As pointed out in the introduction, instabilities in nonlinear
periodic
   structures are a topic of intense theoretical and experimental
   interest and have been discussed in many publications to date.
   In particular, the exact nature of the instabilities, in
   particular in 3-D, is still a matter of controversy. In this
   work, we limit ourselves to comparing our experimental results
   to a simple 1-D numerical simulation. A more in-depth
   discussion of 3-D simulations and analytical calculations will
   be presented elsewhere.

   Figure~\ref{instab_theo} shows the results of a numerical
integration
   of the one-dimensional Gross-Pitaevskii equation with the
   parameters of our experiment. The visibility was calculated in
   the same way as was done for the experimental interference
   patterns. It is clear from this simulation that it is, indeed,
   the nonlinearity that is responsible for the instability at the
   edge of the Brillouin zone. The nonlinearity can be characterized
through the
   parameter~\cite{choi99}
\begin{equation}
C=\frac{\pi n_0 a_s}{k_L^2},
\end{equation}
where $n_0$ is the density of the condensate and
$a_s=5.4\,\mathrm{nm}$ is the $s$-wave scattering length of
$^{87}Rb$. For the experimental parameters used in the experiment
described here, the value of $C$ was $\approx 0.008$.

   When $C$ is set to $0$ in the numerical simulation, the
   visibility remains unaltered when the zone edge is crossed,
   whereas for finite values of $C$ the visibility decreases
  as a quasimomentum of $1$ is approached. Furthermore, the larger the
   value of $C$, the more pronounced the decrease in visibility
   near the band edge. For $C=0.008$,
   corresponding to the value realized in our experiment, the
   onset of the instability is located just below a quasimomentum
   of $0.8$. Experimentally, we find that the visibility starts
   decreasing consistently beyond a quasimomentum of $\approx
   0.6-0.7$, agreeing reasonably well with the results of the
   simulation.
   \section{Discussion and outlook}\label{discussion}
   The experimental results presented in this work demonstrate
   that at the edge of the Brillouin zone, a BEC in an optical
   lattice exhibits unstable behaviour. These instabilities are
   reproduced in a 1-D numerical simulation, and our experimental
   findings also agree qualitatively with recent 3-D simulations~\cite{scott03}.
   The fact that we observe a significant effect of the
   instabilities below accelerations of
   $\approx0.5\,\mathrm{m\,s^{-2}}$, for which the condensate
   spends more than $2\,\mathrm{ms}$ in the critical region around
   the edge of the Brillouin zone (having an extension of around
   $1/10$ of the BZ~\cite{wu03}), indicates that the growth rate of
the instability should be of the order of $500\,\mathrm{s^{-1}}$.
   This agrees reasonably well with a rough estimate of $\approx
   300\,\mathrm{s^{-1}}$ derived from a recent work by Wu and
   Niu~\cite{wu03}.

   Clearly, our approach has a number of shortcomings that prevent
   us from achieving a more quantitative agreement with theory and
   a better characterization of the instabilities. Firstly, in our
   experiment the condensate exhibits a residual sloshing motion,
   meaning that its velocity is not always exactly zero when we
   load it into the optical lattice. This results in an
   uncertainty in the quasimomentum which we estimate to be of the
   order of $10-15\%$ and leads to a considerable scatter in the
   experimental data both of the visibility and the radial width,
   possibly masking a sharper transition between the stable and
   unstable regions. Also, the fact that our trap is almost
   isotropic makes it difficult for us to increase the nonlinear
   parameter $C$ through the condensate density by increasing the
   trap frequency, as this would result in an even stronger
   restoring force along the lattice direction, making it impossible
to cross the BZ edge with a constant (and small)
   acceleration. In order to overcome this problem, it would be
   advantageous to use, for instance, a dipole trap which allows
   one to increase the radial trapping frequency whilst still
   maintaining a small longitudinal frequency. Another interesting
question to address is whether the fact that the direction of our
   optical lattice is at a small angle (a few degrees) to the axis of
the magnetic trap might lead to chaotic motion, contributing to the
   increase in radial width of the interference pattern we
observe~\cite{fromhold03}.

 In summary, the present experimental
observations confirm that Bose-Einstein condensates may be used to
simulate  a variety of nonlinear physics configurations that are
also of interest to other areas of physics, such as solid state
physics. Bose-Einstein condensates offer the added advantage of a
large flexibility in the parameters and in the exploration of
phenomena occurring on a millisecond time scale.
\section*{Acknowledgments}
We gratefully acknowledge stimulating discussion with Robin Scott
and Andrew Martin.  This work was supported by the MURST (PRIN2000
Initiative), the INFM (PRA `Photonmatter'), and by the EU through
Contract No. HPRN-CT-2000-00125.

   \end{document}